\documentclass[usegraphicx,usenatbib]{mn2e}
\usepackage{url}

\bibpunct{(}{)}{;}{a}{}{,}
\usepackage{aasmacros}
\usepackage{color}

\newcommand{\eg}{e.g.,\ }
\newcommand{\ie}{i.e.,\ }
\newcommand{\Msun}{M_{\odot}}

\newcommand{\Lsun}{L_{\odot}}

\newcommand{\Nifs}{$^{56}$Ni}

\def\lsim{\mathrel{\rlap{\lower 4pt \hbox{\hskip 1pt $\sim$}}\raise 1pt\hbox {$<$}}}
\def\gsim{\mathrel{\rlap{\lower 4pt \hbox{\hskip 1pt $\sim$}}\raise 1pt\hbox {$>$}}}

\title[Detectability of High-Redshift Superluminous Supernovae - II. Beyond z=6]
{Detectability of High-Redshift Superluminous Supernovae
with Upcoming Optical and Near-Infrared Surveys - II. Beyond z=6}

\author[Tanaka et al.]{
\parbox[t]{\textwidth}{
Masaomi Tanaka$^{1}$\thanks{E-mail: masaomi.tanaka@nao.ac.jp},
Takashi J. Moriya$^{2,3,4}$, and
Naoki Yoshida$^{5,3}$
}
\vspace*{6pt}\\
$^1${\it National Astronomical Observatory of Japan, Mitaka, Tokyo 181-8588, Japan} \\
$^2${\it Department of Astronomy, Graduate School of Science, University of Tokyo, 7-3-1 Hongo, Bunkyo-ku, Tokyo 113-0033, Japan}\\
$^3${\it 
Kavli Institute for the Physics and Mathematics of the Universe (WPI), Todai Institutes for Advanced Study, the University of Tokyo,}\\
{\it \ \ Kashiwa, Chiba 277-8583, Japan}\\
$^4${Research Center for the Early Universe, Graduate School of Science, University of Tokyo, 7-3-1 Hongo, Bunkyo-ku, Tokyo 113-0033, Japan}\\
$^5${\it Department of Physics, Graduate School of Science, University of Tokyo, 7-3-1 Hongo, Bunkyo-ku, Tokyo 113-0033, Japan}\\
}
\date{Accepted ---. Received ---}

\volume{000}

\pagerange{1\,-\,??} \pubyear{---}

\begin{document}
\maketitle

\label{firstpage}

\begin{abstract}
Observational identification of the first stars is 
one of the great challenges in the modern astronomy.
Although a single first star is too faint to be detected,
supernova explosions of the first stars can be bright enough.
An important question is whether such supernovae 
can be detected in the limited observational area
with realistic observational resources.
We perform detailed simulations to study the detectability of superluminous supernovae (SLSNe) at high redshifts, using the observationally-calibrated 
star formation rate density and supernova occurrence rate.
We show that a 100 deg$^2$ survey with the limiting magnitude of 26 mag 
in near-infrared wavelengths
will be able to discover about 10 SLSNe at $z>10$.
If the survey is extended to 200 deg$^2$ with 27 mag depth,
about 10 SLSNe can be discovered at $z > 15$.
We emphasize that the observations 
at $\ge$ 3 $\mu$m are crucial to detect and select 
SLSNe at $z > 10$.
Our simulations are also applied to the planned survey
with Euclid, WFIRST, and WISH. 
These surveys will be able to detect about 1000, 400, and 3000 SLSNe up to 
$z \sim$ 5, 7, and 12, respectively.
We conclude that detection of SLSNe at $z>10$ is 
in fact achievable in the near future.
\end{abstract}



\section{Introduction}

The first stars, or Population III stars, are 
predicted to be formed at redshift $z \gsim 15$ 
in the standard cold dark-matter scenario
\citep[\eg][]{bromm99,abel00,yoshida03,oshea07,turk09,stacy10,bromm09,bromm11,greif11,clark11}.
Observational identification of the first stars
is one of the great challenges in the modern astronomy.
In fact, a cluster of first stars can be detected in the future
with the next-generation telescopes, such as 
James Webb Space Telescope
(JWST, see \eg \citealt{bromm01,gardner06}).
However, an isolated, single first star is too faint to be observed 
\citep{gardner06,rydberg13}.

Detecting supernova (SN) explosions of the first stars is
an interesting possibility worth pursuing.
A single SN explosion can give rise 
to a luminosity of $L \gsim 10^{9}\ \Lsun$, 
powered by the radioactive energy or the kinetic energy of the explosion.
Note that the luminosity of SNe is comparable to
high-redshift galaxies recently discovered with Hubble Space Telescope
(HST, \eg \citealt{bouwens11}).
This is more than a few orders of magnitude higher 
than the luminosity of a single very massive star 
($L \sim 10^7\ \Lsun$ for a 500 $\Msun$ star, \citealt{bromm01}).

Because of this advantage, 
detectability of SNe at high redshift
have been studied in the past literature
\citep[\eg][]{miralda-escude97,mesinger06,whalen13CCSN}.
These studies revealed, however, that 
normal core-collapse SNe are too faint to be detected at $z>6$.
Detection of normal SNe at $z > 6$
requires observations deeper than 30 AB mag 
in near-infrared (NIR) wavelengths,
which can be reached only with a long exposure of JWST
($\sim 2 \times 10^4$ seconds for 5 $\sigma$ significance).

In this circumstance, several literatures 
\citep{scannapieco05,pan12,pan12b,hummel12,whalen12PISN,whalen13,desouza13} have examined the detectability of
pair-instability SNe (PISNe, see \eg \citealt{heger02,kasen11,dessart13}).
PISNe give rise to an extremely high luminosity ($L \sim 10^{10}\ \Lsun$)
powered by $>$ 1-10 $\Msun$ of \Nifs,
and, they can be as bright as $\sim 26-27$ AB mag in NIR
at $z \gsim 10$ \citep{whalen12PISN,whalen13}.
These works in part were motivated by the discovery of 
a PISN candidate SN 2007bi \citep{gal-yam0907bi,young10}.
It is noted that there is ongoing debate on typical masses of 
Population III stars.
Theoretically, both very massive stars
($\gsim 100 \Msun$, as massive as progenitor of PISNe, 
\eg \citealt{bromm99,abel00}) 
and ordinary massive stars
($\lsim 50 \Msun$, \eg \citealt{yoshida07,hosokawa11}) 
might have existed in the early universe.
Observationally, the existence of ordinary massive stars are
corroborated by chemical abundances of metal poor stars 
(\eg \citealt{frebel09}, 
see \citealt{ren12} for ongoing search of PISN signature).

The recent discovery of superluminous supernovae (SLSNe)
\citep[see][and references therein]{quimby11,gal-yam12}
opens a new window to observe SN explosions of the first stars.
SLSNe are thought to be powered by 
a huge amount of \Nifs\ \citep{umeda08,young10,moriya10} 
and/or strong interaction with the circumstellar material (CSM).
The latter is supported by 
characteristic Type IIn spectra, \eg narrow hydrogen emission lines, 
in some SLSNe-II (SLSN with hydrogen,
according to the classification by \citealt{gal-yam12}),
such as SNe 2006gy \citep{ofek07,smith0706gy,smith0806gy,agnoletto09,kawabata09}
and 2008fz \citep{drake10}.
Numerical simulations of radiation hydrodynamics
of SN explosion with dense CSM
have been performed by \citet[hereafter M13]{moriya13}, 
who found that their models can reproduce the light curve of SN 2006gy.
Thanks to the high luminosity, 
SLSNe are ideal targets at high redshift Universe.
In fact, a few SLSNe and luminous Type IIn SNe 
have been detected at $z \sim 2-4$
\citep{cooke09,cooke12}.

It has been argued by \citet{cooke08,quimby11}; M13; \citet{whalen13IIn} that 
SLSNe and luminous Type IIn SNe are bright enough to be detected at $z > 4$.
However, SLSNe are known to be extremely rare 
($\sim 10^{-3}$ of core-collapse SNe, \citealt{quimby11,gal-yam12,quimby13}).
Thus, an important question still remains;
is there enough number of high-redshift SNe 
in the limited observational area that can be observed 
with realistic observational resources?

\begin{figure}
  \includegraphics[scale=1.3]{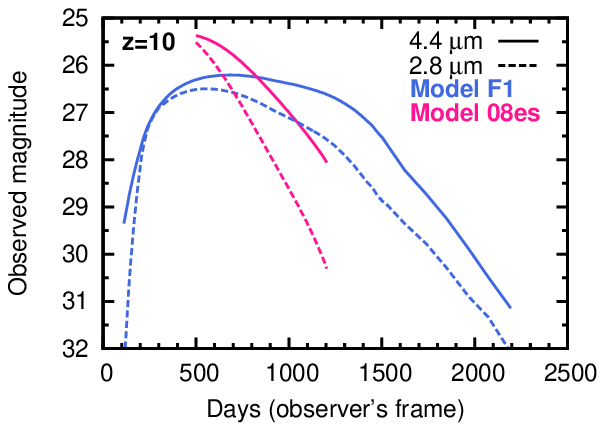} 
  \includegraphics[scale=1.3]{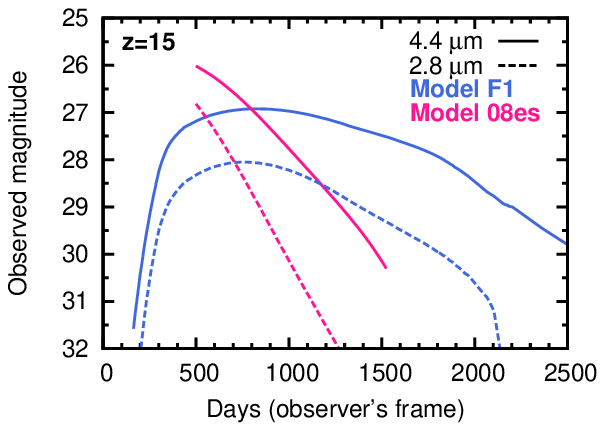} 
  \includegraphics[scale=1.3]{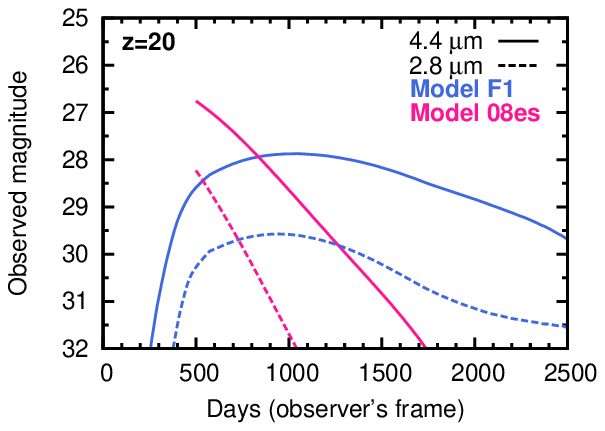} 
  \caption{Observed-frame light curves of Model F1 (blue) and SN 2008es (pink) 
at redshift $z =$ 10 (top), 15 (middle), and 20 (bottom). 
The solid and dashed lines
represent the magnitude in 4.4 $\mu$m and 2.8 $\mu$m, respectively.}
\label{fig:LCobs}
\end{figure}

Earlier in our paper \citep[hereafter Paper I]{tanaka12SLSN},
we studied the detectability of high-redshift SLSNe 
in the limited observational area for the first time.
We showed about 100 SLSNe up to $z \sim 4$ can be detected
with upcoming Subaru/Hyper Suprime-Cam (HSC) Deep survey,
which reaches 24.5 mag depth in $z$-band.
By using Ultra Deep survey for 3.5 deg$^2$ (25.6 mag in $z$-band),  
the maximum redshift can be as high as $z \sim 5$.
We also showed that deep NIR survey can detect SLSNe even at $z \sim 6$.

In the present paper, we extend the study of Paper I to redshifts 
beyond $z=6$, which is near the end of reionization of the Universe,
with a special emphasis on $z > 10$, the era of the first star formation.
We first describe our models for SLSNe in Section \ref{sec:model}.
The method and setup for mock observations are described in
Section \ref{sec:method}.
Results of simulations are presented in Section \ref{sec:results}.
Based on the results, the optimized survey strategy is proposed in 
Section \ref{sec:optimization}.
We discuss selection methods to pick up 
high-redshift SLSNe in Section \ref{sec:selection}.
Then, we apply our simulations to planned NIR surveys 
in Section \ref{sec:application}.
Finally, we give conclusions in Section \ref{sec:conclusions}.

Throughout the paper, 
we assume the $\Omega_M = 0.3$, $\Omega_{\Lambda}=0.7$
and $H_0 = 70$ ${\rm km\ s^{-1}\ Mpc^{-1}}$ cosmology.
The magnitudes are given in the AB magnitude
unless otherwise specified.


\section{Models of Superluminous Supernovae}
\label{sec:model}

\begin{figure*}
  \begin{tabular}{cc}
  \includegraphics[scale=1.3]{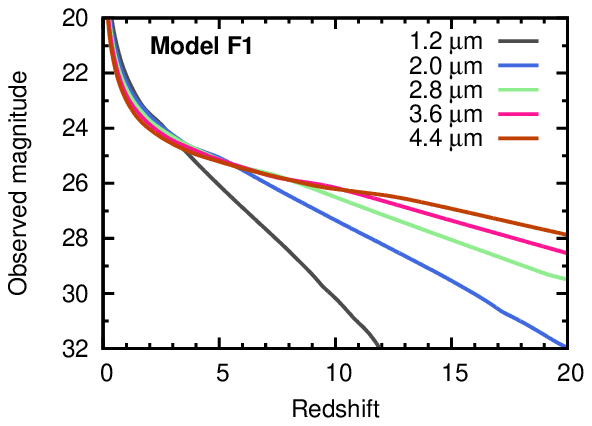} 
  \includegraphics[scale=1.3]{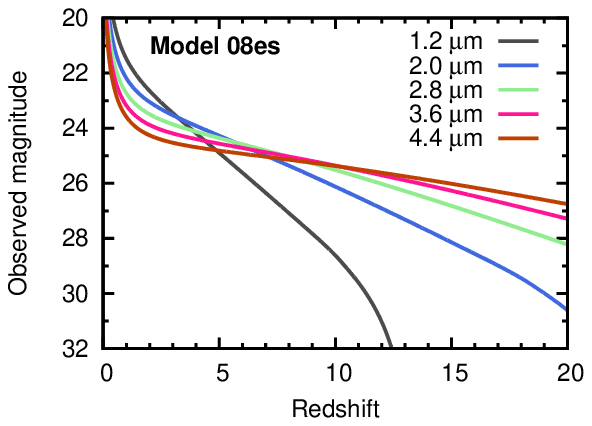} 
  \end{tabular}
  \caption{Observed peak magnitude of Model F1 (left) and SN 2008es (right)
as a function of redshift. 
To detect SNe at $z>10$, the observations at $>$ 3 $\mu$m are crucial.
A typical limiting magnitude per visit with the planned wide-field NIR satellites 
is 24-26 mag (see Section \ref{sec:application}).
The faintest limit of each panel (32 mag) corresponds to 
$\sim 10^6$ s integration with JWST.}
\label{fig:LCpeak}
\end{figure*}

To study the detectability of SLSNe at high redshifts,
we use two models for spectral evolution of SLSNe.
One is a theoretical model of interacting SNe 
and the other is actual observational results.
For a large part of the simulations shown in this paper, 
we use the latter, observationally-calibrated model.

For the theoretical model, we adopt
the result of radiation hydrodynamic simulations by M13.
We use their "Model F1".
In this model, the SN ejecta has mass 
of $M = 20\ \Msun$ and kinetic energy of $E = 1 \times 10^{52}$ erg.
This SN ejecta collides with the dense CSM with $M = 15\ \Msun$.
The density distribution of the CSM is
assumed to be flat ($\rho (r)$ = constant),
and the inner and outer radius of the CSM is set to be
5 $\times 10^{15}$ cm and 11 $\times 10^{15}$ cm, respectively.
Details are found in M13.

This model nicely reproduces the light curve of SN 2006gy 
\citep{ofek07,smith0706gy,smith0806gy,agnoletto09,kawabata09}.
The peak absolute magnitude is about $-21.5$ mag in the optical wavelengths,
and the duration around the peak is about 100 days.
The color temperature reaches $T \sim 15,000$ K near the peak.
The behavior of the light curves depends on the 
distribution and the mass of the dense CSM.
Note that although the formation scenario of the dense CSM 
might depend on 
the metallicity of the progenitor, we keep using the same model
over the wide range of redshifts.

For this model, we construct apparent-magnitude light curves
for various redshifts.
By taking $L_{\nu}$ spectra from the simulations,
we numerically compute K-correction by using Equation 8 of \citet{hogg02}.
For the bandpass filters, we adopt the broad-band filter 
of JWST/NIRCAM
\footnote{\url{http://www.stsci.edu/jwst/instruments/nircam/instrumentdesign/filters/index_html}}.
In Figure \ref{fig:LCobs}, we show the light curves of
Model F1 at redshift $z=10$, 15, and 20 (from top to bottom).
In the observer's frame, the timescale around the peak 
is as long as 1000 days at $z > 10$.

Figure \ref{fig:LCpeak} shows the peak magnitudes of the Model F1
as a function of redshifts.
At higher redshifts, the brightness in the shorter wavelengths 
becomes dramatically fainter.
Thus, the observations at $\ge$ 3 $\mu$m are crucial
in order to detect SNe at $z > 10$.
Note that a typical limiting magnitude per visit with the planned 
wide-field NIR satellites is 24-26 mag (see Section \ref{sec:application}).
The faintest limit of each panel (32 mag) corresponds to 
$\sim 10^6$ s integration with JWST.

The other model is based on the actual observation of one of the SLSNe,
SN 2008es \citep{miller09,gezari09}.
SN 2008es is a SLSN with the weak hydrogen lines,
and it was classified as Type IIL SN.
The absolute peak magnitude of SN 2008es ($\sim -22$ mag in optical) 
is brighter than that of SN 2006gy,
while the decline rate is faster than SN 2006gy.
We select SN 2008es as the second model because
SN 2008es has characteristics in contrast to SN 2006gy,
\ie a higher luminosity and a faster timescale.
Such different behaviors may result from
different distribution and/or mass of the CSM \citep[\eg][]{moriya12}
or completely different explosion mechanisms \citep[\eg][]{kasen10}.

For spectral evolution of SN 2008es,
we adopt the blackbody model by \citet{miller09}.
The observations of SN 2008es are available 
only after the maximum brightness.
The blackbody temperature evolves from 15000 K to 6000 K (at 60 days).
The blackbody fit serves as an approximation of the observed
spectrum. The flux at short wavelengths is somewhat uncertain
owing to the limited bandpass filters.
Although there is a known variety of SEDs in SLSNe,
the effective blackbody temperature and its evolution
are not quite different among SLSNe 
(\citealt{quimby11,chomiuk11}, see also Paper I).
Hereafter we call this model "Model 08es".

We show the light curves of SN 2008es at $z>10$ in Figure \ref{fig:LCobs},
which shows a clear contrast to Model F1.
The peak magnitude of Model 08es is brighter than Model F1
by about 1 mag.
Figure \ref{fig:LCpeak} shows the peak magnitudes as a function of 
redshifts.
As in Model F1, the wavelengths longer than 3 $\mu$m progressively become 
more important at higher redshifts.


\section{Method and Setup for Mock Observations}
\label{sec:method}

\subsection{Method of Simulations}

We generate SNe for given survey parameters
and perform mock observations of generated SNe.
The method of the simulation is similar to that in Paper I.
We briefly describe the method here.

We first setup the redshift grids at $z=0-20$, 
with the interval of $dz = 0.01$.
In each redshift bin, the number of SNe is computed 
according to the adopted SN rate (see Section \ref{sec:rate}).

The multi-band light curves of the SLSN models (see Section \ref{sec:model}) 
in the observer's frame are computed in each redshift bin.
To account for the observed dispersion in the peak luminosity of SLSNe 
\citep[see][]{gal-yam12},
we introduce a dispersion of $\sigma = 0.3$ mag.
We adopt this relatively small value
so that the bright end of the luminosity function 
does not affect the detectability at high redshifts.

The effect of extinction is crudely included
when we use Model 08es 
because the extinction in the host galaxy is {\it not} 
corrected in the model by \citet{miller09}.
Note that the host galaxies of SLSNe are underluminous
\citep{quimby07,neill11}, 
and the typical host extinction seems to be small.
At higher redshifts, the intergalactic absorption by 
neutral hydrogen is not negligible.
To take this absorption into account, 
we set the model flux below Lyman limit to be zero.

The generated model SNe are observed 
with a certain observational strategy.
We set (1) the duration of a survey, 
(2) the frequency to visit the same field (or cadence),
(3) the survey area, 
and (4) the detection limit per visit. 
Note that the detection limit per visit does not necessarily match
the limiting magnitude of each observation.
The limiting magnitude can also be that 
of the stacked images for a period within the cadence.
Considering the light curves in Figure \ref{fig:LCobs}, 
we keep 3-year survey duration 
and 3-month cadence unless otherwise mentioned.
We explore a variety of the survey area and detection limit.

In this paper, we focus on SNe at $z>6$.
Thus, the observation in NIR wavelengths is a natural choice.
We first assume simultaneous 
observations at 1-5 $\mu$m with 5 broad-band filters
in Sections \ref{sec:results} and \ref{sec:optimization}.
Then, we also perform simulations with planned sets of 
filters in Section \ref{sec:application}.
As shown in Figure \ref{fig:LCpeak}, 
we note that
observations around 1 $\mu$m are not essential to detect high-redshift SNe.
For the bandpass filters, we adopt those of JWST/NIRCAM;
F115W, F200W, F277W, F356W, and F444W 
although the observations with JWST are not necessarily assumed.
For the detection limit, 
we assume the same limiting magnitude in these 5 filters.
We do not consider the effect of host galaxy contamination 
to the limiting magnitudes.

We impose stringent detection criteria as in Paper I.
Our definition of detection is 
fulfilling {\it both} of the following two criteria;
(1) brighter than the detection limit 
in more than 2 bands at least at one epoch,
{\it and} (2) brighter than the limiting magnitude
at more than 3 epochs at least in one band.

\begin{figure}
  \includegraphics[scale=1.3]{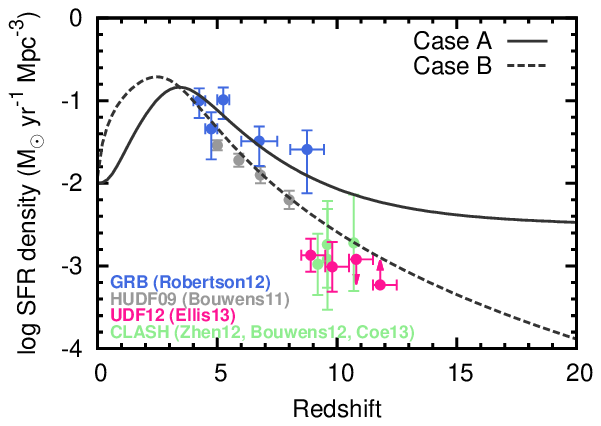} 
  \caption{
The SFR density estimated by galaxies
\citep{bouwens11,bouwens11apj,zheng12,bouwens12,coe13,ellis13}
and GRBs \citep{robertson12}. 
We test two cases of the SFR density (solid and dashed lines for Cases A and B,
respectively).
Case A is the SFR density model by \citet{robertson12}, which is the
lower bound of the SFR density derived from GRBs.
Case B is a simple extrapolation of the formula by \citet{hopkins06},
which is consistent with the galaxy measurements.
}
\label{fig:SFR}
\end{figure}

\begin{figure*}
  \begin{tabular}{cc}
  \includegraphics[scale=1.3]{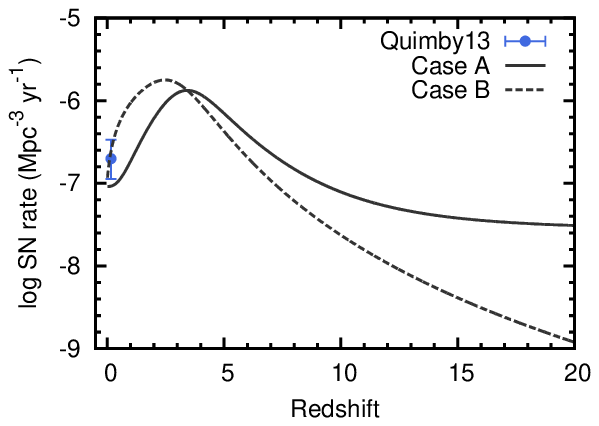} 
  \includegraphics[scale=1.3]{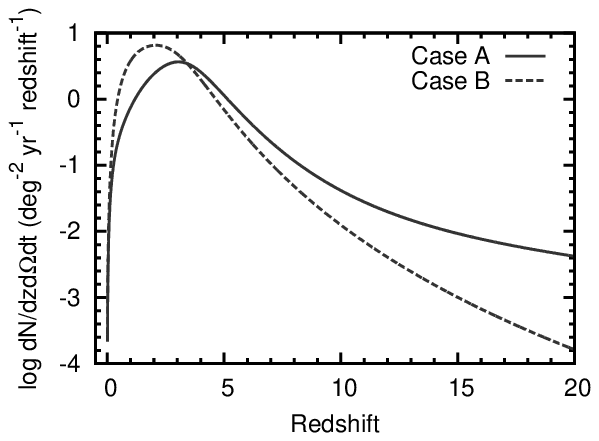} 
  \end{tabular}
  \caption{
(Left) The cosmic occurrence rate of SLSNe per unit volume.
The solid and dashed lines represent Cases A and B SFR density models.
Blue point is the observed total SLSN rate by \citet{quimby13}.
(Right) SLSN rate per unit area and per unit time in the {\it observer's frame}.}
\label{fig:SNR}
\end{figure*}

\subsection{Rate of Superluminous Supernovae}
\label{sec:rate}

A key ingredient of our simulations is 
cosmic occurrence rate of SLSNe at high redshifts.
It is reasonable to assume that the occurrence rate
is proportional to the cosmic star formation rate (SFR) density $\rho_{*}$.
We adopt the SFR density derived from existing observations.
Figure \ref{fig:SFR} shows various measurements of SFR density
using galaxies \citep{bouwens11,bouwens11apj,zheng12,bouwens12,coe13,ellis13}
and gamma-ray bursts (GRBs, \citealt{ishida11,robertson12}).
To cover the possible range of the SFR density,
we adopt two cases; (A) SFR density model by \citet{robertson12},
which is consistent with the lower bound of the 
SFR density derived from GRBs (solid line in Figure \ref{fig:SFR}),
and (B) model by \citet{hopkins06} extended to $z=20$ (dashed line).

Since the progenitors of SLSNe are thought to be massive stars,
the delay time is negligible. 
With this assumption, the rate of SLSNe ($R_{\rm SLSN}$) can be written 
using $\rho_{*}$ as
\begin{equation}
R_{\rm SLSN}(z) 
= f_{\rm SLSN} \ \rho_{*}(z) 
\frac{ \int_{M_{\rm min, SLSN}}^{M_{\rm max, SLSN}} \psi(M) {\rm d}M }{ \int_{M_{\rm min}}^{M_{\rm max}} M \psi(M) {\rm d}M },
\label{eq:rate}
\end{equation}
where $\psi(M)$ is the stellar initial mass function 
(IMF, $ \psi(M) \propto M^{-(\Gamma +1)} $).
We adopt a modified Salpeter A IMF of \citet{baldry03}
with the slope $\Gamma=0.5$ for $0.1\Msun\ (=M_{\rm min}) < M < 0.5 \Msun$ and
$\Gamma=1.35$ for $0.5\Msun < M < 100 \Msun \ (=M_{\rm max})$.

As in Paper I, we assume that 
(1) from the wide mass range of stars,
only massive stars with the mass of 
$M_{\rm max, SLSN}=50 \Msun$ - $M_{\rm max, SLSN}=100 \Msun$
can be a potential progenitor of SLSNe, and 
(2) a fraction $f_{\rm SLSN}$ of such massive stars actually 
explode as SLSNe
\footnote{Since at least one progenitor of Type IIn SN is
known to be as massive as 
$M_{\rm ZAMS} > 50-80 \Msun$ \citep{gal-yam07,gal-yam09},
we set the minimum mass of SLSNe to be $M_{\rm max, SLSN}=50 \Msun$.}.
The fraction $f_{\rm SLSN}$ can be calibrated by the 
observational constraints of the SLSN rate.
\citet{quimby13} estimated the rate of SLSNe
to be $2 \times 10^{-7}\ {\rm Mpc^{-3} yr^{-1}}$
at $z \sim 0.2$
\footnote{Although \citet{quimby13} derived the rates of 
SLSN-I (SLSN without hydrogen, $3 \times 10^{-8}\ {\rm Mpc^{-3} yr^{-1}}$)
and SLSN-II (SLSN with hydrogen, $1.5 \times 10^{-7}\ {\rm Mpc^{-3} yr^{-1}}$) 
separately, we simply use the total rate.}.
Adopting the SFR density from \citet{hopkins06}, 
this rate is obtained if $f_{\rm SLSN}$ is set to be $2 \times 10^{-2}$
for $M_{\rm max, SLSN}=50 \Msun$.
This fraction corresponds to $10^{-3}$ of total core-collapse SNe 
(with the progenitor mass range of $M = 8-100 \Msun$).
Hereafter we use this value of $f_{\rm SLSN}$ for both Cases A and B SFR density.
Since redshift evolution of this fraction is poorly 
understood both observationally and theoretically, 
$f_{\rm SLSN}$ is assumed to be constant over redshifts.
Possible impact of different IMFs is briefly discussed 
in Section \ref{sec:application} (see also Paper I).
Note that Paper I adopted $f_{\rm SLSN} = 2 \times 10^{-3} - 2 \times 10^{-2}$,
which gave conservative estimates.

The left panel of Figure \ref{fig:SNR} shows 
the SN rate per unit comoving volume as a function of redshift.
The solid and dashed lines represent the SN rate 
with Cases A and B SFR density, respectively.
The SN rate with Case B SFR density is 
consistent with the observed rate of SLSNe 
(blue point, \citealt{quimby13}).
The adopted SN rates are also roughly consistent with the rate
$\sim 4 \times 10^{-7}\ {\rm Mpc^{-3} \ yr^{-1}}$ 
derived using a single detection at $z \sim$ 2 and 4 
by \citet{cooke12} although this rate may not represent 
the total SLSN rate.
The right panel of Figure \ref{fig:SNR} shows 
the SN rate per unit area of the sky per redshift and per
unit time in the observer's frame.
The expected number of SNe is an order of 0.01 
${\rm deg^{-2} \ yr^{-1}\ redshift^{-1}}$ at $z > 10$.

\section{Results}
\label{sec:results}

\begin{figure}
  \includegraphics[scale=1.3]{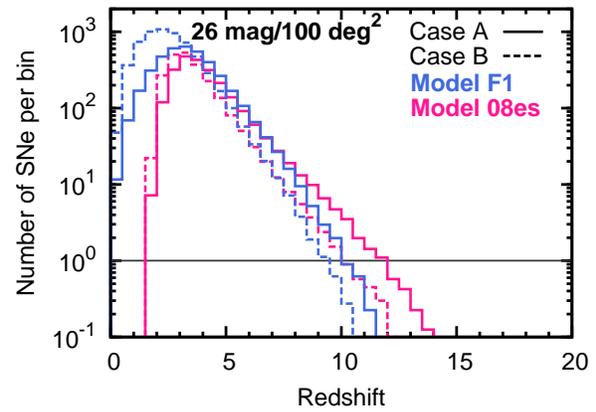} 
  \caption{Expected number of SN detection per $dz=0.5$ bin as a function of redshift
with the survey area of 100 ${\rm deg^{2}}$
and the limiting magnitude of 26 mag in 1-5 $\mu$m.
The solid and dashed lines show the dependence on the adopted SFR density
(Cases A and B, respectively).
The blue and red lines represent Model F1 and Model 08es, respectively.}
\label{fig:hist_100deg2_26mag}
\end{figure}

\begin{figure}
  \includegraphics[scale=1.3]{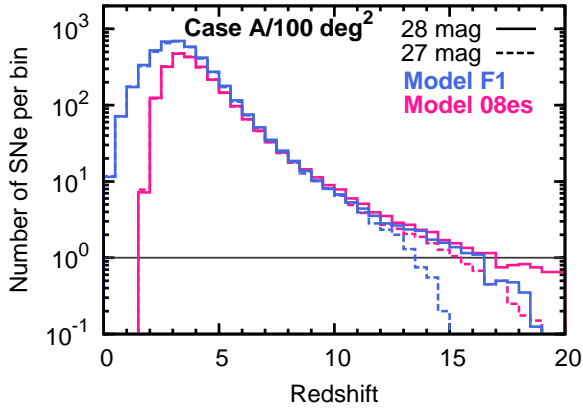} 
  \caption{Expected number of SN detection per $dz=0.5$ bin as a function of redshift
with the survey area of 100 ${\rm deg^{2}}$
and the limiting magnitude of 27 mag (solid line) and 
28 mag (dashed line) in 1-5 $\mu$m.
Case A SFR density is adopted.
The blue and red lines represent Model F1 and Model 08es, respectively.}
\label{fig:hist_100deg2_28mag}
\end{figure}

We first show the results of our fiducial survey;
observations with the survey area of 100 ${\rm deg^{2}}$, 
and the limiting magnitude of 26 mag.
The survey duration and cadence are set to be 3 years and 3 months,
respectively.
Figure \ref{fig:hist_100deg2_26mag} shows the 
expected number of SNe per $dz=0.5$ bin as a function of redshift.
The expected number of SNe at $z > 10$ is an order of 1-10
with Model 08es (red lines in Figure \ref{fig:hist_100deg2_26mag}).

The solid and dashed lines show the dependence on the adopted 
SFR density (Cases A and B, respectively).
The expected number with the Case A SFR density is 
higher than that with Case B by a factor of about 3 at $z>6$,
as expected from Figures \ref{fig:SFR} and \ref{fig:SNR}.

Figure \ref{fig:hist_100deg2_28mag} shows the 
dependence on the limiting magnitudes.
The observations deeper than 27 mag at $> 3 \mu$m
are deep enough not to miss SLSNe at $z \sim 15$ 
(see also Figure \ref{fig:LCpeak}).
With 100 ${\rm deg^2}$ area, SLSNe beyond $z=15$ can be detected.

\begin{figure}
  \includegraphics[scale=1.3]{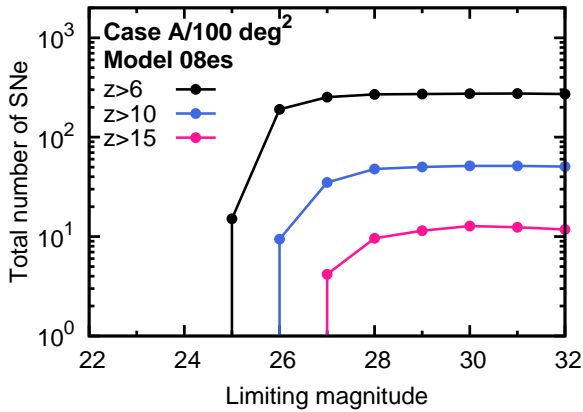} 
  \caption{Expected total number of 
SLSNe beyond $z=6$ (black), 10 (blue), and 15 (red)
as a function of limiting magnitude.
Case A SFR density and Model 08es are adopted.}
\label{fig:num_mag}
\end{figure}

Figure \ref{fig:num_mag} summarizes the expected number of 
SLSNe beyond $z=6$ (black), 10 (blue), and 15 (red)
as a function of limiting magnitude
for the case of 100 ${\rm deg^{2}}$ survey area.
The Case A SFR density and Model 08es are adopted.
This figure leads us to conclude the following.
(1) To detect SLSNe at $z>6$, 10, and 15, 
observations deeper than 25, 26, and 27 mag are required, respectively.
(2) Observations deeper than 28 mag are not needed
to increase the number of SLSNe.
Especially, the latter suggests that 
{\it in order to discover high-redshift SLSNe efficiently,
observational resources should be devoted to enlarge the survey area,
instead of making the observation deeper than 28 mag.}
The optimized survey strategy is discussed in the next section.

\section{Optimized survey strategy}
\label{sec:optimization}

\begin{figure*}
  \begin{tabular}{cc}
  \includegraphics[scale=1.4]{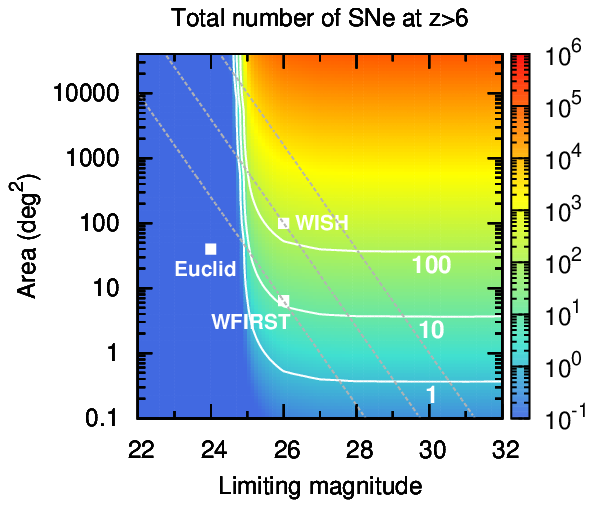} \\
  \includegraphics[scale=1.4]{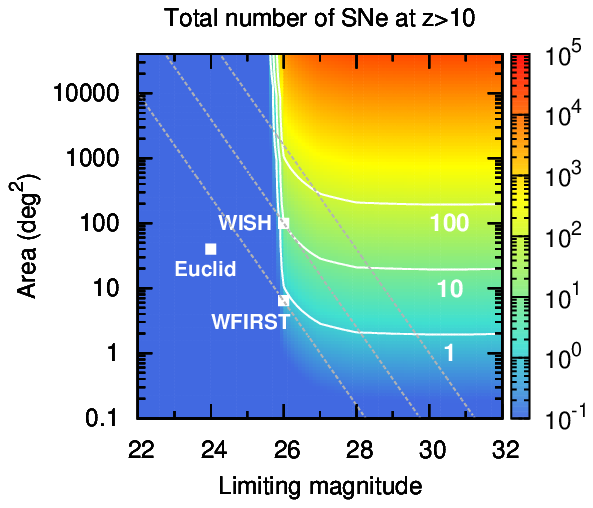} 
  \includegraphics[scale=1.4]{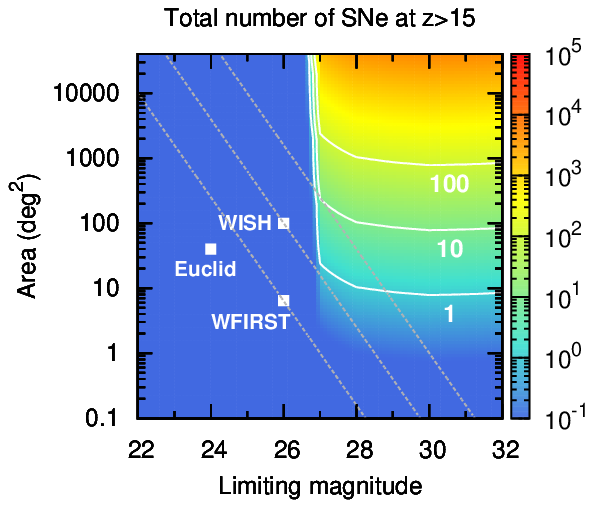} 
  \end{tabular}
  \caption{
Expected total number of SLSNe at $z>6$ (upper), 10 (lower left) and 15 (lower right)
as a function of survey area and limiting magnitude.
The contours show the combination of the survey area and limiting magnitude
giving 1, 10, and 100 SLSNe (from left bottom to right top).
White squares show the survey area and limiting magnitude 
for planned surveys (Table \ref{tab:param}).
The contours are nearly vertical at the limiting magnitudes of 25 mag ($z>6$),
26 mag ($z>10$), and 27 mag ($z>15$).
This indicates that the observations deeper than at least 
25, 26 and 27 mag are required to detect SLSNe at $z>6$, 10 and 15, respectively.
On the other hand, the contours are horizontal at the limiting magnitude 
deeper than 28 mag.
This indicates that, for a given survey area, 
observations deeper than 28 mag do not increase 
the number of SLSNe dramatically.
The gray dashed lines represent 
the combination of the survey area and depth for a given 
survey power (\ie $A\Omega t$).
Case A SFR density and Model 08es are adopted.}
\label{fig:depth_area}
\end{figure*}

Since observational resources are limited, 
it is important to find an optimized survey strategy.
In general, survey observations can be characterized by 
(1) survey area and (2) survey depth.
For the case of transient survey, 
(3) cadence is another important factor.
However, SLSNe at high redshifts have a long timescale (Figure \ref{fig:LCobs}),
and thus, the requirement for the cadence is not strong
(we fix the cadence to be 3 months in all the simulations 
presented so far).
Thus, optimization of the survey strategy means 
finding the best combination of the survey area and depth
to maximize the number of SLSNe.

Figure \ref{fig:depth_area} shows the expected total number of SLSNe 
at $z > 6$ (upper), 10 (lower left) and 15 (lower right) 
in the two-dimensional plane 
of the survey area and depth
(see Appendix for similar simulations for optical surveys).
The contours show the survey area and depth giving 
1, 10, and 100 SLSNe (from left bottom to right top).
Figure \ref{fig:depth_area} includes the survey area up to 
the whole sky.
Note that, in actual survey, the visibility of the sky 
that can be visited multiple times during the survey period is limited.
When observations with a NIR satellite are considered, 
a typical maximum area with nearly permanent visibility
is about 1000 deg$^2$ around the ecliptic poles.

The contours are nearly vertical around 25 mag ($z>6$),
26 mag ($z > 10$), and 27 mag ($z>15$).
This clearly indicates that, irrespective of the survey area, 
observations at least deeper than 25 mag, 26 mag and 27 mag
are required to detect SLSNe at $z>6$, 10, and 15, respectively
(see also Figure \ref{fig:num_mag}).
On the other hand, 
the contour becomes horizontal at limiting magnitudes deeper than 28 mag.
This indicates that, for a given survey area, 
observations deeper than 28 mag do not increase the number of SNe.

The gray dashed lines in Figure \ref{fig:depth_area} represent
the combination of the survey area and depth for a given 
survey power (\ie $A\Omega t$, a product of certain set of 
photon-correction power $A$, field of view $\Omega$, 
and observational time $t$).
From the comparison with simulations,
we conclude that 
the optimized survey strategy to detect more than 10 SLSNe at $z>10$ 
would be {\it 100 deg$^2$ survey with the limiting magnitude of 26 mag}.

In order to detect more than 10 SLSNe at $z>15$,
wider {\it and} deeper survey should be performed.
The optimized strategy 
would be {\it 200 deg$^2$ survey with the limiting magnitude of 27 mag}.
Compared with the survey to detect 10 SLSNe at $z > 10$,
the survey power $A\Omega t$ should be increased by a factor of $\sim 13$.

Our results can also be applied for PISNe.
The existence or occurrence rate of PISNe are not established well,
and the predicted rates have a wide range:
$10^{-2} - 1 \ {\rm deg^{-2}\ yr^{-1}\ redshift^{-1}}$ 
\citep{wise05,weinmann05,johnson13}.
Note that these rates may increase by the effect of rotation 
of progenitor stars \citep{chatzopoulos12}.
When the prediction by \citet{johnson13} is adopted, 
the rate is $10^{-2} - 10^{-1} \ {\rm deg^{-2}\ yr^{-1}\ redshift^{-1}}$,
which is similar to the expected SLSN rate (Figure \ref{fig:SNR}).
According to the calculations by \citet{whalen12PISN,whalen13},
the peak brightness of some PISN models is as bright as 
26 mag at $z=10$ and 27 mag at $z=15$ in the NIR wavelengths, 
which are also similar to those of SLSNe 
(Figures \ref{fig:LCobs} and \ref{fig:LCpeak}).
In this case, the same observing strategy with SLSNe will be able to discover
a similar number of PISNe:
a 100 deg$^2$ survey with the limiting magnitude of 26 mag will discover
about 10 PISNe at $z>10$, and 
a 200 deg$^2$ survey with the limiting magnitude of 27 mag will discover
about 10 PISNe at $z>15$.

\section{Selection of High-Redshift SLSNe}
\label{sec:selection}

\begin{figure*}
  \begin{tabular}{cc}
  \includegraphics[scale=1.2]{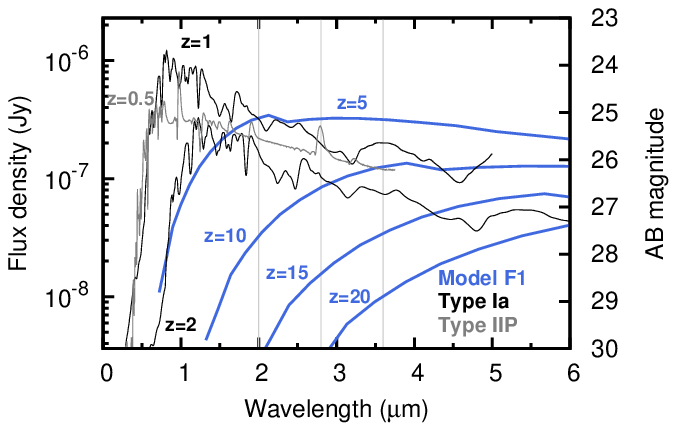} 
  \includegraphics[scale=1.2]{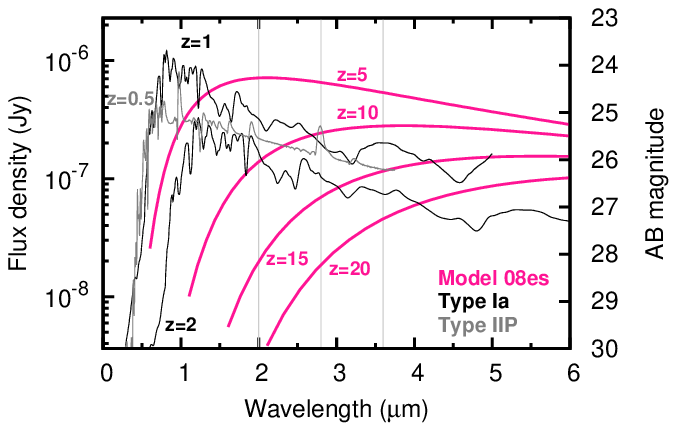} 
  \end{tabular}
  \caption{SEDs of
Model F1 (left) and Model 08es (right) for different redshifts.
For comparison, spectral templates of Type Ia SNe (black) 
and Type IIP SNe (gray) are also shown.}
\label{fig:SED}
\end{figure*}

\begin{figure}
  \includegraphics[scale=1.7]{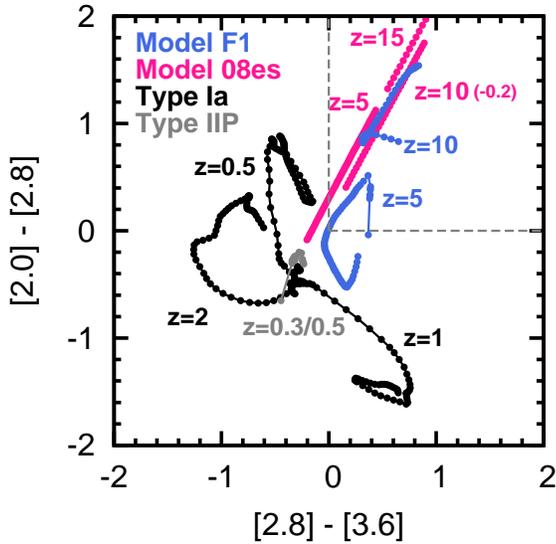} 
  \caption{Color-color diagram for two SLSN models (blue and pink).
They are compared with Type Ia SNe and Type IIP SNe at lower redshifts
with similar observed magnitudes.
SLSNe at high redshifts have a red color 
both in [2.0]$-$[2.8] and [2.8]$-$[3.6] 
compared with Type Ia and IIP SNe with similar observed magnitudes.
Colors of Model 08es at $z=10$ are shifted by $-0.2$ mag 
in [2.0]$-$[2.8] color just for visibility.}
\label{fig:color}
\end{figure}

We discuss selection methods for SLSNe at high redshifts.
Since SLSNe are rare objects, it is naturally expected that
more Type Ia SNe and normal core-collapse SNe at lower redshifts will be discovered 
with the NIR surveys presented in the previous sections.
Thus, we must efficiently pick up candidates of high-redshift SLSNe.

A clear difference is its timescale of the variability.
Because of the intrinsically long timescale and time dilation by high redshifts,
the expected timescale of the variation of SLSNe is longer than 100 days
(Figure \ref{fig:LCobs}).
This is much longer than that of Type Ia SNe at lower redshifts ($z \lsim 2$).

However, Type IIP SNe may have a similar timescale at the plateau phase.
For the further selection, observations more than 2 bands are helpful.
Figure \ref{fig:SED} shows the spectral energy distribution (SED) of high-redshift SLSNe,
compared with those of low-redshift Type Ia SNe (black) and Type IIP SNe (gray)
with similar observed magnitudes.
For the SEDs of Type Ia and IIP SNe, we use spectral templates by 
\citet{nugent02} \footnote{\url{http://supernova.lbl.gov/~nugent/nugent_templates.html}}.
In the NIR wavelengths, SLSNe at high redshifts are observed
around the peak of the SED or at the bluer side of the peak.
In contrast, Type Ia and IIP SNe are always observed at 
the redder side of the peak.
As a result, SLSNe at high redshifts are redder 
than Type Ia and IIP SNe with similar observed magnitudes.

The red color of high-redshift SLSNe is more clearly seen 
in a color-color diagram (Figure \ref{fig:color}).
In this figure, we use bandpass filter of JWST 
([2.0], [2.8], and [3.6] are magnitudes in 
F200W, F277W, and F356W filters, respectively).
SLSNe at high redshifts tend have a redder color 
in [2.0]$-$[2.8] and [2.8]$-$[3.6],
compared with Type Ia and IIP SNe with similar observed magnitudes.
We note, however, that only with 2-band observations, 
confusion with Type Ia SNe are not fully solved.
We emphasize that observations at $\ge$ 3 $\mu$m
is useful not only for detection at higher redshifts,
but also for the target selection.
Objects having both [2.0]$-$[2.8] $>0$ and [2.8]$-$[3.6] $>0$ colors
(dashed lines in Figure \ref{fig:color})
are likely to be high-redshift SLSNe.

\begin{table*}
\begin{center}
\caption{Parameters for upcoming NIR surveys}
\label{tab:param}
\begin{tabular}{cccccc}
\hline
\noalign{\vspace{2pt}}
Survey           & Area       &  Depth  &  Wavelength range   &   Cadence  &  Duration  \\
                 & (deg$^2$)   &  (mag) &                    &          &   \\
\noalign{\vspace{2pt}}
\hline\hline
Optimal ($z>10$)      &   100     &  26.0                       &   3-5$\mu$m       &   3 months  &  3 years \\
Optimal ($z>15$)      &   200     &  26.0                       &   3-5$\mu$m       &   3 months  &  3 years \\
\hline
Euclid           &   40     &  24.5 (visual), 24.0($Y, J, K$)   &  0.55 - 2 $\mu$m (visual - $H$) &   10 days  &  3 years \\
WFIRST           &   6.5    &  26.0                             &  0.73 - 2.4 $\mu$m ($Z$ - $K$)  &   5 days  &  1.8 years \\
WFIRST-extended           &   100    &  26.0                         &  0.73 - 2.4 $\mu$m ($Z$ - $K$)  &   5 days  &  1.8 years \\
WFIRST-extended +3$\mu$m  &   100    &  26.0                         &  0.73 - 3.0 $\mu$m              &   5 days  &  1.8 years \\
WISH            &   100    &  26.0                              &  1.0 - 4.5 $\mu$m              &   10 days  &  1 year \\
\hline
\end{tabular}\\
\end{center}
$^{}*$ Limiting magnitude per visit, but it can also be the limiting
magnitude in the stacked image for a given period within the cadence.
\end{table*}

\section{Application to Future Surveys}
\label{sec:application}

\begin{figure}
  \includegraphics[scale=1.4]{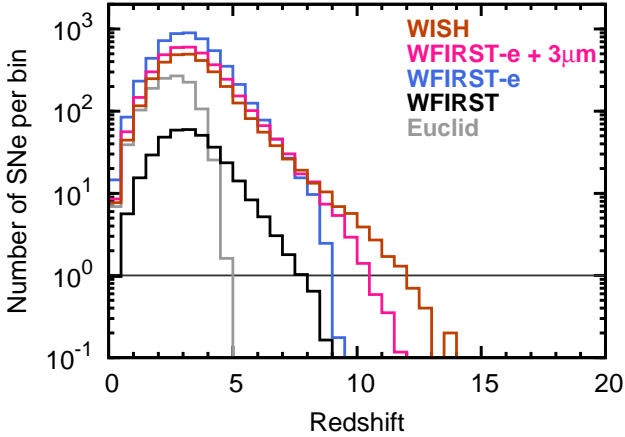}   
  \caption{Expected number of SN detection per $dz=0.5$ bin as a function of redshift
with upcoming NIR surveys. 
For the adopted survey parameters, see Table \ref{tab:param}.
WFIRST (black and blue lines) covers the wavelength range up to
$K$-band (2.4 $\mu$m). If a 3 $\mu$m-band is hypothetically added,
the highest redshift will exceed $z=10$ (pink line).
Since WISH plans to cover up to 4$\mu$m, SNe at higher redshifts
can be detected. 
All the simulations have been performed for Case A SFR density and Model 08es.
}
\label{fig:hist_casestudy}
\end{figure}

Based on the results shown in the previous sections,
we apply our simulations to several planned surveys in the near future.
We consider NIR survey with 
Euclid\footnote{\url{http://sci.esa.int/euclid}},
the Wide-Field Infrared Survey Telescope (WFIRST)
\footnote{\url{http://wfirst.gsfc.nasa.gov}},
and Wide-field Imaging Surveyor for High-redshift (WISH)
\footnote{\url{http://www.wishmission.org/en/index.html}}.
Adopted survey parameters are summarized in Table \ref{tab:param}.

Euclid plans to perform a wide (15000-20000 deg$^2$) and deep (40 deg$^2$) survey 
\citep{laureijs11}.
A part of deep survey may have multiple visits 
for Type Ia SNe, although the exact cadence is not yet fixed. 
The wavelength coverage is up to $2 \mu$m ($H$ band).
The survey depth per visit is 24.5 mag in the visual band and 24.0 mag in the NIR bands.
We hypothetically assume 10-day cadence and 3-year survey period.
We perform mock observations with these parameters.
Throughout this section, we adopt Case A SFR density and Model 08es.
The result is shown in the gray line in Figure \ref{fig:hist_casestudy}. 
Euclid will be able to discover $\sim 1000$ SLSNe at $z<5$.
The maximum reachable redshift is about $z=5$, 
which is limited by the 24.0 mag depth at 2 $\mu$m 
(see Figure \ref{fig:LCpeak}).

Euclid will provide an unprecedented sample of SLSNe.
Since about half of SLSNe detected by Euclid are located 
at $z < 2$, they can also be observed with optical spectroscopy
\citep{cooke09,cooke12}.
With this sample, we can study statistical properties 
of each class of SLSNe (SLSN-I and SLSN-II, see \citealt{quimby11,gal-yam12}),
such as the luminosity function 
and statistics of the light curve duration, 
which can be used to unveil the nature and progenitors of SLSNe.

Next, we consider surveys with WFIRST.
WFIRST plans to perform dedicated SN survey in a part of 
the observational time \citep{green12}.
The planned survey area is 6.5 deg$^2$ (wide) and 1.8 deg$^2$ (deep).
Since a large survey area is critical for the detection of SLSNe 
(Figure \ref{fig:depth_area}), 
we adopt 6.5 deg$^2$.
The wavelength coverage is up to 2.4 $\mu$m ($K$-band)
and the survey depth per visit is 26.0 mag.
Cadence and survey period are 5 days and 1.8 years, respectively,
which are optimized for Type Ia SNe.
The result of mock observation with these parameters 
(WFIRST, Table \ref{tab:param})
is shown in black line in Figure \ref{fig:hist_casestudy}.
Thanks to the deep observations at 2.4 $\mu$m, 
WFIRST will be able to discover $\sim 400$ SLSNe up to $z \sim 7$.

It is also shown, however, that the survey area of 6.5 deg$^2$ 
is not enough to fully utilize its observational depth.
To see this effect, we hypothetically perform the simulations
with the survey area of 100 deg$^2$ 
with all the other parameters kept the same 
(WFIRST-extended, Table \ref{tab:param}).
Such a survey dramatically increases the number of SLSNe
(see Figure \ref{fig:depth_area}):
$\sim 6000$ SLSNe up to $z \sim 9$ will be detected
(blue line in Figure \ref{fig:hist_casestudy}).

Even with the extended survey area, there is a clear cutoff in the 
expected number of SLSNe below $z \sim 10$.
This is because 
the expected brightness of SLSNe in the $K$-band (1.8-2.4 $\mu$m) 
becomes dramatically fainter 
at the redshifts higher than $z \sim 5$ (Figure \ref{fig:LCpeak}).
To see the advantage to have 3$\mu$m band,
we perform a simulation by hypothetically 
adding 3$\mu$m band to the WFIRST-extended survey
(pink line in Figure \ref{fig:hist_casestudy}).
If WFIRST possibly covers 3$\mu$m,
it will be able to carry out the nearly ideal survey,
detecting SLSNe at $z>10$.
We emphasize the importance of observations at $\ge$ 3 $\mu$m.

WISH plans to focus on the deep survey with 
$\sim 100$ deg$^2$ \citep{yamada12}.
The wavelength range is 1.0-4.5 $\mu$m
and the survey depth per visit is about 26.0 mag.
Cadence and survey period are not yet fixed, and thus, 
we hypothetically assume 10-day cadence and 1-year survey period
(see WISH in Table \ref{tab:param}).
The brown line in Figure \ref{fig:hist_casestudy} shows the 
expected number of SLSNe with WISH survey.
WISH will be able to detect about 3000 SLSNe in total.
Thanks to the wavelength coverage up to 4.5 $\mu$m,
the maximum redshift is higher than those of Euclid and WFIRST.
It may be able to discover SLSNe up to $z \sim 12$.
This is, in fact, quite similar to the optimized survey strategy 
to detect SLSNe at $z>10$ suggested in Section \ref{sec:optimization}.

WISH and extended WFIRST surveys will be able to detect
more than 100 SLSNe at $z > 6$ (see also Figure \ref{fig:depth_area}).
Such high-redshift SLSNe can be spectroscopically observed with JWST
and also ground-based 30m-class telescopes, such as 
Thirty Meter Telescope (TMT, 
see \eg \citealt{wright10} for the expected sensitivity) 
\footnote{\url{http://www.tmt.org}}, 
Giant Magellan Telescope (GMT) \footnote{\url{http://www.gmto.org}}, and
European Extremely Large Telescope (E-ELT)
\footnote{\url{http://www.eso.org/public/teles-instr/e-elt.html}}.
As demonstrated in Paper I, this number is sensitive to 
the slope of IMF (see also \citealt{cooke09}). 
If the slope $\Gamma$ changes from 1.35 to 1.1, 
the expected number increases by a factor of 3.
Comparison between the redshift evolution of the SLSN rate and SFR density
will provide a unique method to probe top-heavy IMF
at high-redshift Universe.

The planned and model surveys discussed above are not necessarily 
tuned to the detection of high-redshift SLSNe.
Especially, the cadence is higher than that required to detect 
high-redshift SLSNe (see Figure \ref{fig:LCobs}).
By stacking 3-month data of WISH or WFIRST-extended survey, 
the observational depth can be 27 mag in NIR.
This is in fact very close to the ideal survey strategy to 
detect SLSNe at $z \sim 15$ (Section \ref{sec:optimization}).
It is emphasized that the proposed optimized survey strategy 
to detect SLSNe at $z>$10-15 can be realized 
with a slight modification of the planned surveys.

\section{Conclusions}
\label{sec:conclusions}

SN explosions of first stars are unique possibility to 
observationally study a single first star.
We study the detectability of SLSNe at high redshifts, 
including the era of the first star formation.
It has been suggested that 
SLSNe can be bright enough even at such high redshifts, 
but it was not clear if SLSNe can be detected 
for a limited observational area and 
with realistic observing resources.

We perform simulations of mock observations for SLSNe
using the observationally-calibrated 
SFR densities and supernova rates.
We find that a 100 deg$^2$ survey with the limiting magnitude of 26 mag
will be able to discover $\sim 10$ SLSNe at $z > 10$.
To extend the detection to $z > 15$,
the survey should be extended to 200 deg$^2$ with 27 mag depth.
We emphasize that the observations
at $\ge$ 3 $\mu$m are important to detect SLSNe at $z > 10$.
The observations deeper than 28 mag do not increase the 
number of SNe, and observational resources should be devoted
to enlarge the survey area.

High-redshift SLSNe can be distinguished from lower-redshift
Type Ia SNe and normal core-collapse SNe 
by the long timescale of variability.
In addition, the red observed colors are important characteristics
to select SLSNe at high redshifts.
Objects that are red both in [2.0]$-$[2.8] and [2.8]$-$[3.6] color
are likely to be high-redshift SLSNe.
The observations at $\ge$ 3 $\mu$m are also important for 
the target selection.

We also applied our simulations to planned surveys 
with the wide-field NIR satellites.
We find that the survey by Euclid, WFIRST, and WISH
will be able to detect 
about 1000, 400, and 3000 SLSNe up to $z \sim$ 5, 7, and 12, respectively.
It is demonstrated again that the observations at $\ge$ 3 $\mu$m 
is crucial to detect SLSNe at $z>10$.
Among these survey satellites, the observations with WISH seems 
the most suitable to detect high-redshift SLSNe.
By stacking 3-month data, SLSNe even at $z \sim 15$ can be discovered.

We emphasize that the proposed optimized survey strategy 
to detect SLSNe at $z>10$ is not far from reality.
In fact, we show that the planned NIR surveys 
partly achieve the required specification,
and that a slight modification of the planned surveys 
makes the surveys closer to the ideal survey to detect SLSNe at $z>10$.
We will be able to reach a single star at $z>10$,
possibly out of the first stars, 
with such NIR surveys in the near future.
Such surveys will provide a unique way to unveil
the properties of the first stars and IMF in the early Universe.

\vspace{3mm}
\noindent
MT thanks Kohji Tsumura for useful comments.
The authors are supported by the Japan Society 
for the Promotion of Science (JSPS)　
Grant-in-Aid for Young Scientists (24740117: MT, 20674003: NY), 
Grant-in-Aid for Scientific Research (25287050: NY),
and Grant-in-Aid for JSPS Fellows (23.5929: TM).
This research has been supported in part by World Premier
International Research Center Initiative, MEXT, Japan.



\appendix

\renewcommand{\thetable}{A\arabic{table}}

\begin{table*}
\begin{center}
\caption{Parameters for upcoming optical surveys}
\label{tab:param_opt}
\begin{tabular}{ccccc}
\hline
\noalign{\vspace{2pt}}
Survey           & Area       &  Depth    &   Cadence  &  Duration  \\
                 & (deg$^2$)   &  (mag)   &            &   \\
\noalign{\vspace{2pt}}
\hline\hline
HSC Deep $^{*}$       &   30     &  25.0   &   10 days  &  0.5 years \\
HSC UltraDeep $^{*}$  &   3.5    &  26.0   &   10 days  &  0.5 years \\
LSST               &   20000   & $^{**}$  &   10 days  &  0.25 years $\times 10$ \\
LSST deep drilling &   100    &  26.0   &   10 days  &  0.5 years \\
\hline
\end{tabular}\\
\end{center}
$^{*}$ For simplicity, a constant limiting magnitude is assumed. 
See Paper I for more realistic simulations.\\
$^{**}$ Limiting magnitude per single visit for LSST: 
23.9 ($u$), 25.0 ($g$), 24.7 ($r$), 24.0 ($i$), 23.3 ($z$), and 22.1 ($y$).
\end{table*}

\section{Application to Upcoming Optical Surveys}
\label{sec:opt}

\begin{figure*}
  \begin{tabular}{cc}
  \includegraphics[scale=1.4]{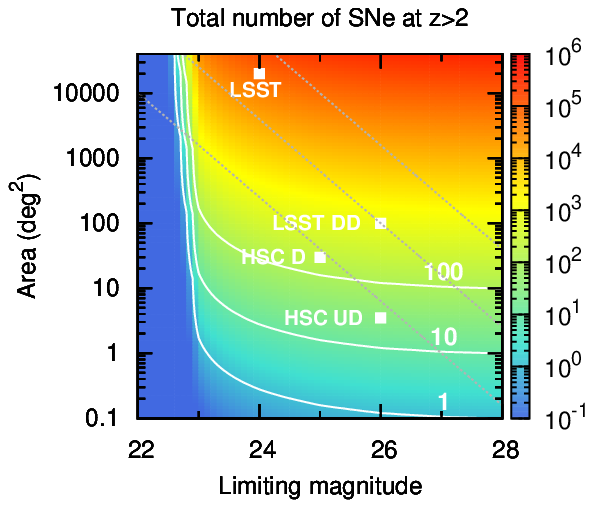}   
  \includegraphics[scale=1.4]{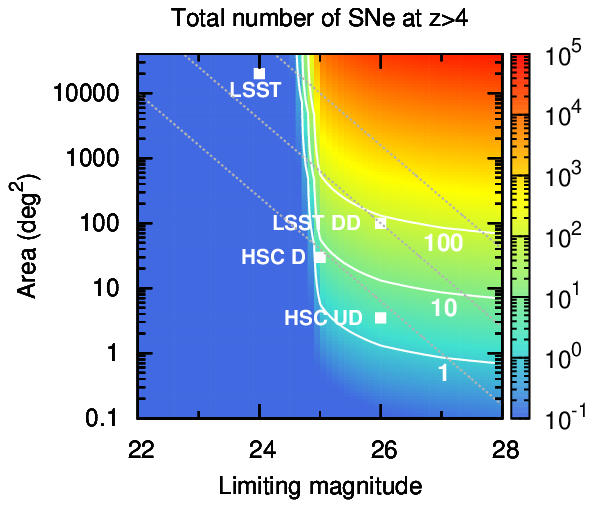}   
  \end{tabular}
  \caption{
The same as Figure \ref{fig:depth_area} 
but for optical surveys with 10-day cadence for 0.5 years.
The left and right panels show 
the expected total number of SLSNe at $z>2$ and 4, respectively.
White squares show the survey area and limiting magnitude 
for planned optical surveys (Table \ref{tab:param_opt}).
Case B SFR density and Model 08es are adopted.}
\label{fig:area_depth_opt}
\end{figure*}

\begin{figure}
  \includegraphics[scale=1.4]{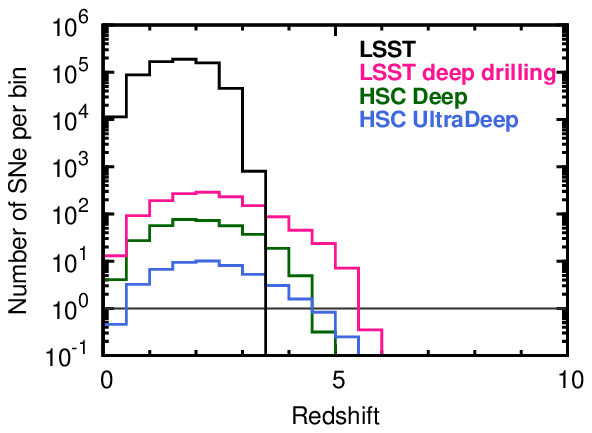}   
  \caption{Expected number of SN detection per $dz=0.5$ bin as a function of redshift
with LSST surveys. 
For the adopted survey parameters, see Table \ref{tab:param_opt}.
For comparison, we show the simulations with the HSC survey strategy
(Deep and UltraDeep layers).
In this figure, a constant limiting magnitude in optical wavelengths
is assume for simplicity.
See Paper I for more detailed simulations with HSC.
All the simulations have been performed for Case B SFR density and Model 08es.
}
\label{fig:hist_opt}
\end{figure}

We also apply our simulations to upcoming optical surveys.
Paper I performed detailed simulations with
realistic, planned observational strategy 
for Subaru Hyper Suprime-Cam (HSC, \citealt{miyazaki06}) survey.
On the other hand, as shown in Section \ref{sec:optimization}, 
it is useful to study a wide range of parameters 
to find the optimized survey strategy.
In this section, we present simpler simulations for optical surveys,
but with a wider parameter space.

With optical surveys, detection of SLSNe up to $z \sim 5$ 
is expected (Paper I).
Thus, it is natural to adopt a shorter duration of the survey 
and a higher cadence than those of NIR surveys.
Here we simply adopt
(1) 0.5-year survey and (2) 10-day cadence.
For simplicity, simultaneous observations in the optical $ugrizy$ bands 
with the same limiting magnitudes are assumed.
Figure \ref{fig:area_depth_opt} shows the expected total number of 
SLSNe at $z>2$ (left) and 4 (right) as a function of 
survey area and limiting magnitude per visit.
All the simulations have been performed for Case B SFR density
(better calibrated at $z<6$ than Case A SFR density) 
and Model 08es.

Our simulations are also applied for planned survey with
Large Synoptic Survey Telescope (LSST, \citealt{ivezic08,lsst09}).
LSST will perform 20000 deg$^2$ survey in the optical $ugrizy$ bands.
Each visit consists of a short exposure (15 seconds), 
giving following limiting magnitudes;
23.9 ($u$), 25.0 ($g$), 24.7 ($r$), 24.0 ($i$), 23.3 ($z$), and 22.1 ($y$).
Each patch of the sky is visited about 100-200 times for 10 years.
Thus, the parameters of the survey using the single visit 
can be roughly approximated as (1) 3-month survey for each year 
(repeating for 10 years) and (2) 10-day cadence (see Table \ref{tab:param_opt}).
The black line in Figure \ref{fig:hist_opt} shows the expected number 
of SLSNe using the single visits of LSST.
LSST will be able to discover about $10^6$ of SLSNe up to $z \sim 3-4$.
In this overwhelming sample, some SLSNe ($\gsim 10-100$) may be 
discovered as extremely luminous sources because of 
the magnification by gravitational lensing (\citealt{oguri10},
see also \citealt{takahashi11}).

By appropriately stacking the data, giving a deeper limiting magnitude,
the maximum redshift will be higher.
One promising strategy is using "deep drilling" fields of LSST.
About 10 \% of the observing time will be devoted to 
observe several deep fields.
Even with 10 \% of the observing time, 
if we limit the survey area to 100 deg$^2$
and the survey duration to 0.5 year,
deep observations (with about 100 images stacked)
can be performed with 10-day cadence.
Such a deep image will give limiting magnitudes of about 26 mag
(Table \ref{tab:param_opt}, LSST deep drilling).
Simulations with these parameters show that 
LSST deep drilling observations will be able to detect 
about $10^3$ SLSNe up to $z \sim 5-6$
(pink line in Figure \ref{fig:hist_opt}).

\end{document}